\documentclass{article}

\usepackage[preprint]{neurips_2026}

\usepackage[utf8]{inputenc}
\usepackage[T1]{fontenc}
\usepackage{hyperref}
\usepackage{url}
\usepackage{booktabs}
\usepackage{amsfonts}
\usepackage{nicefrac}
\usepackage{microtype}
\usepackage{xcolor}
\usepackage{graphicx}
\usepackage{subcaption}
\usepackage{multirow}
\usepackage{tikz}
\usepackage{pifont}
\usepackage{wrapfig}
\usepackage{listings}
\newcommand{\squishlist}{
   \begin{list}{$\bullet$}
    { \setlength{\itemsep}{0pt}      \setlength{\parsep}{3pt}
      \setlength{\topsep}{3pt}       \setlength{\partopsep}{0pt}
      \setlength{\leftmargin}{2em} \setlength{\labelwidth}{1em}
      \setlength{\labelsep}{0.5em} } }
\newcommand{\squishend}{
    \end{list}  }
\newcommand{\squishenum}{
   \begin{list}{\arabic{enumi}.}
    { \usecounter{enumi}
      \setlength{\itemsep}{0pt}      \setlength{\parsep}{3pt}
      \setlength{\topsep}{3pt}       \setlength{\partopsep}{0pt}
      \setlength{\leftmargin}{1.0em} \setlength{\labelwidth}{1em}
      \setlength{\labelsep}{0.5em} } }
\newcommand{\squishenumend}{
    \end{list}  }

\newcommand{\system}{Slipstream}

\title{\system{}: Trajectory-Grounded Compaction Validation for Long-Horizon Agents}

\author{
  \textbf{Zhuofu Chen}$
  \quad
  \textbf{Rui Pan}$
  \quad
  \textbf{Yinwei Dai}$
  \quad
  \textbf{Ravi Netravali}$
  \\
  Princeton University\\
    \texttt{\{zhuofuc,ruipan,yinweid,rnetravali\}@princeton.edu}
}

\begin{document}

\maketitle

\begin{abstract}

To cope with the large contexts that long-horizon LLM agents produce, modern frameworks increasingly rely on compaction -- invoking an LLM to rewrite the accumulated trajectory into a shorter summary that the agent resumes from. Today, compaction runs synchronously on the critical path of agent execution but this can unpredictably degrade accuracy due to a structural validation gap: the compactor must condense context but is fundamentally unaware of precisely what information the agent will need later. Further, because post-compaction agent steps are conditioned on the new summary, targeted validation criteria do not exist and errors silently propagate through coherent but incorrect behavior. Our key insight is that asynchronous compaction efficiently addresses this gap: by running the compactor in parallel with continued agent execution on the original context, the candidate summary and the agent's next steps are generated independently from the same pre-compaction state, yielding a validation signal independent of the summary itself. We build \emph{\system{}}, a trajectory-grounded compaction system that uses a judge to validate the candidate summary against the agent's continued reasoning, checking that it preserves both the agent's forward intent and the key facts and constraints it depends on. Across long-horizon coding (SWE-bench Verified) and web-browsing (BrowseComp) workloads, \system{} improves task accuracy by up to \textbf{8.8 percentage points} while reducing end-to-end latency by up to \textbf{39.7\%}.
We open source \system{} at \url{https://github.com/chenzhuofu/slipstream}.

\end{abstract}

\vspace{-10pt}
\section{Introduction}
\label{sec:introduction}
\vspace{-5pt}

Long-horizon LLM agents have rapidly become a dominant paradigm for solving complex tasks that require sustained reasoning over many rounds of tool use, observation, and planning~\citep{react,sweagent,toolformer,webgpt}. As these agents tackle longer horizons -- from multi-step software engineering to extended web research -- their context windows accumulate observations, intermediate results, and partial progress that may become relevant many steps later. This growth quickly degrades model behavior: accuracy drops with context length even when all relevant information remains present, a phenomenon known as context rot with reported accuracy drops of 14–85\%~\citep{lost_in_middle, context_rot, context_length_alone_hurts}. To manage this, modern agent frameworks have converged on \textit{compaction}: invoking an LLM to rewrite the accumulated trajectory into a shorter summary that the agent then resumes from directly~\citep{anthropic_compaction, microsoft_compaction, langmem_summarization}. Compaction is now triggered routinely throughout long-horizon execution rather than as a last resort at the model's context limit -- Anthropic, for example, recommends compacting at as little as 5–20k tokens for some workloads~\citep{anthropic_auto_compaction_cookbook}.

As deployed today, compaction runs \emph{synchronously} on the critical path of agent execution, pausing the agent until the compactor produces a summary of the context. Beyond the direct latency overheads that this introduces -- inflating end-to-end agent execution times by 26--44\% across mainstream workloads -- the deeper issue is one of accuracy. Compaction must decide what to preserve without knowing what the agent will need next, and synchronous execution provides no way to validate this decision: once the summary replaces the original context, every subsequent step is conditioned on the summary itself, so the agent's continued behavior cannot serve as an independent check. When the compactor drops or distorts information the agent later needs, task outcomes silently degrade, with no error signal at the moment of compaction. Traditional summarization metrics do not address this since they operate only on the current context and the candidate summary, not on the future agent behavior that determines whether the summary is sufficient. What is missing is a validation signal grounded in the agent's continued trajectory itself.

Our key insight is that \emph{asynchronous compaction} is the only execution mode that produces this missing validation signal without duplicating the agent’s execution. When the compactor runs in parallel with the agent’s continued execution on the original, uncompacted context, the agent’s next steps and the candidate summary are generated independently from the same pre-compaction input. This independence is crucial: in synchronous execution, any agent behavior used for validation is already conditioned on the summary it is meant to check, removing its value as an independent signal. Asynchrony is therefore not merely a way to remove blocking latency -- it is what makes trajectory-grounded validation possible.

\begin{figure*}[t]
    \centering
    \includegraphics[width=0.87\textwidth]{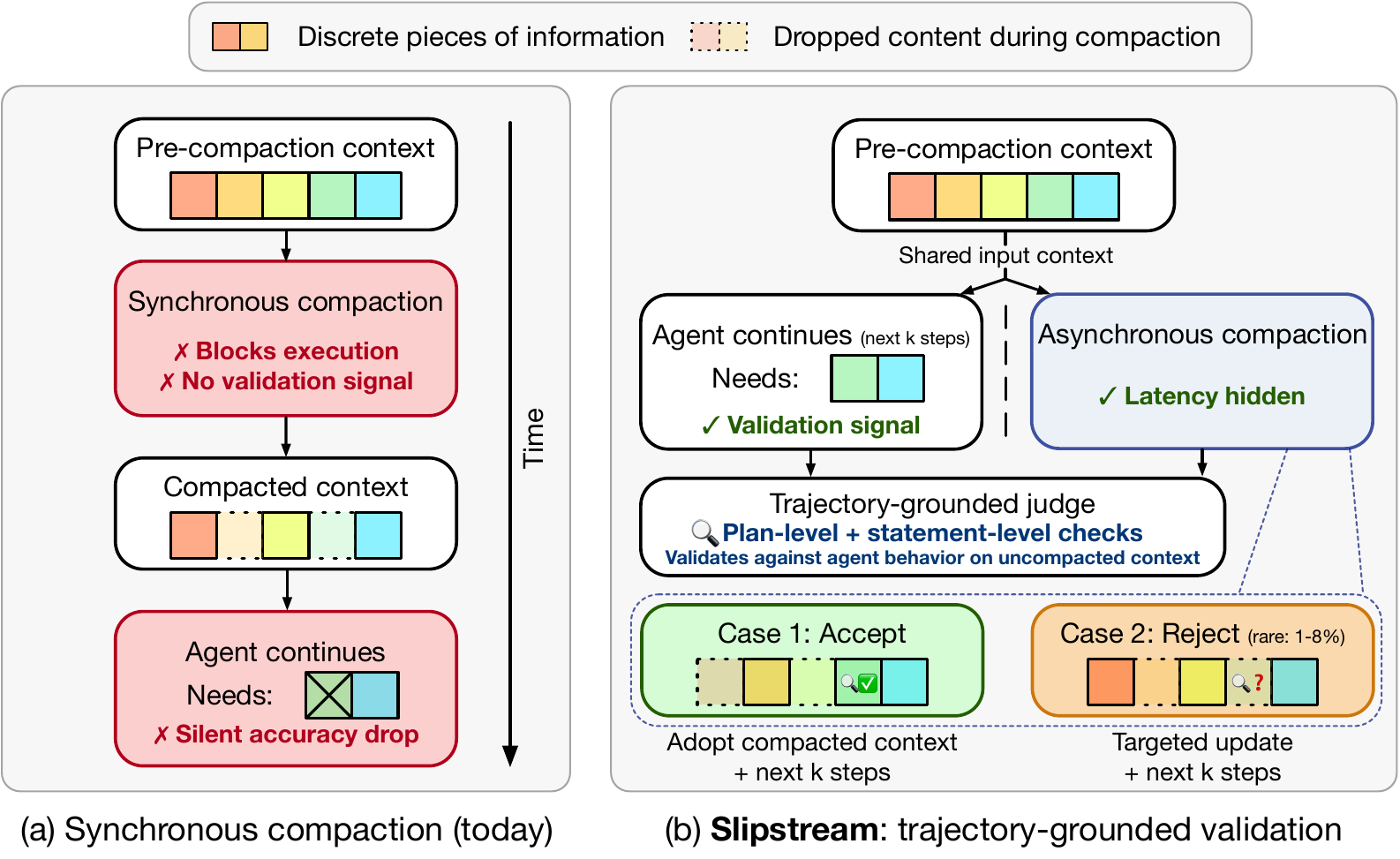}
    \vspace{-5pt}
    \captionsetup{font=small}
    \caption{Synchronous compaction vs. \textit{\system{}}. (a) Synchronous compaction blocks agent execution and offers no visibility into what future actions require, leading to silent accuracy degradation. (b) \textit{\system{}} runs the compactor in parallel with continued agent execution on the original context, hiding compaction latency. The next-$k$ agent actions provide a held-out validation signal covering intent and facts for the trajectory-grounded judge, which either accepts the candidate summary for adoption or triggers a targeted update before adoption.}
    \label{fig:overview}
    \vspace{-20pt}
\end{figure*}

Based on this insight, we present \textbf{\textit{\system{}}}, a trajectory-grounded compaction system for long-horizon agents that runs the compactor in parallel with continued agent execution on the uncompacted context (Figure~\ref{fig:overview}). Once the compactor returns, \system{} applies a judge to validate the candidate summary against the next-$k$ agent steps that completed during compaction. The judge performs two complementary checks, exploiting the structure of agentic execution: a \emph{plan-level} check that the summary preserves the agent's forward intent, and a \emph{statement-level} check that the specific facts, constraints, and intermediate results the agent relies on are preserved. On acceptance (the predominant case),  execution resumes from the compacted state augmented with the already-generated next-$k$ steps; otherwise, \system{} performs a targeted update to the summary using the judge's diagnosis. Two properties make this design feasible. First, compaction errors that affect downstream behavior most often surface within the asynchronous window; across our workloads, 88-100\% of first error manifestations arise within 3 agent steps. Second, long-horizon agent serving is memory-constrained, leaving headroom to run the compactor and judge alongside the main agent at low cost.

Across long-horizon coding (SWE-bench Verified) and web-browsing (BrowseComp) workloads with Qwen3.5-9B and Seed-OSS-36B-Instruct, \system{} improves task accuracy by up to \textbf{8.8 percentage points} over synchronous compaction while reducing end-to-end latency by up to \textbf{39.7\%}. Both wins stem from the same mechanism: asynchrony moves compaction off the critical path, and provides the independent behavioral signal that exposes summary errors synchronous baselines silently adopt. %

\section{Background and Related Work}
\label{sec:background}

Modern agents solve long-horizon tasks by iterating through a reason-then-act loop: they reason over the accumulated conversation, issue tool calls, observe results, and append each step back into the context~\citep{react,sweagent,toolformer,webgpt}. This context is the agent's working state, recording observations, constraints, partial progress, and unresolved subgoals. The rate of context growth varies by agent paradigm -- single-agent strategies accumulate the full trajectory, while task-decomposition approaches~\citep{context_folding, llmcompiler} split work across sub-agents that return shorter results. Regardless, as task horizons grow, context grows until the agent either reaches the model's maximum context length or suffers degraded performance before that limit. This degradation is not merely a capacity issue: LLM accuracy drops with context length even when all relevant information remains present, a phenomenon known as context rot~\citep{lost_in_middle, context_rot, context_length_alone_hurts}, with reported degradations of 13.9--85~\%~\citep{context_length_alone_hurts}. Long-horizon agents thus require active \emph{context management} well before the nominal window is exhausted.

The design space of context management spans three broad classes. First, \emph{KV-cache compression} operates at the inference level, evicting or compressing key-value entries to reduce memory and bandwidth costs while preserving the underlying token sequence~\citep{snapkv, pyramidkv, rocketkv}. These techniques target the computational cost of long contexts but do not change what the model attends to logically, as the agent's working state retains the full trajectory. Second, \emph{retrieval-augmented and external-memory approaches} offload accumulated history to an external store and fetch relevant content on demand, ranging from semantic-similarity retrieval over chunked logs~\citep{rag_lewis} to memory-augmented architectures that maintain evolving structured state~\citep{memgpt, memex, amem}. These methods preserve information losslessly in storage but introduce retrieval-time decisions about what to surface, which can be brittle with noisy or similar fragments~\citep{memex}. Third, \emph{context compaction} rewrites the in-context working state into a shorter representation that the agent continues from directly~\citep{resum, supo, anthropic_context_engineering_cookbook}. Compaction differs from the first two in that it modifies the agent's logical context rather than its physical representation or its access path; the agent operates on the compacted state as if it were the trajectory itself.

Compaction has emerged as the dominant approach for long-horizon agents in recent agent frameworks~\citep{anthropic_context_engineering_cookbook, langmem_summarization}. Earlier heuristic pruning schemes~\citep{sweagent,memgpt,openhands} -- e.g., dropping oldest messages, obsolete tool outputs, or intermediate reasoning traces -- assumed that removable content can be identified by shallow signals like recency or message type, an assumption that is brittle for long-horizon tasks where seemingly incidental information may later become necessary. Compaction, by contrast, uses an LLM to produce a holistic summary that can in principle preserve information based on its semantic relevance rather than its position in the trace. In practice, agent systems trigger compaction aggressively, well before the context is full, to stay ahead of context rot. Anthropic, for example, recommends not waiting until the context is full: compaction should be triggered at roughly 5--20k tokens for some workloads and 50--100k tokens for more complex tasks~\citep{anthropic_auto_compaction_cookbook}. Thus, compaction is now a routine part of long-horizon execution rather than a last-resort recovery mechanism at the context limit.

Today, compaction is typically performed \emph{synchronously}. When triggered, the agent pauses, a compactor LLM generates a summary, the summary replaces the original context, and agent execution resumes. This design is natural for autoregressive generation; the next agent step is conditioned on the current context, so the system updates the context before resuming.
\vspace{-10pt}

\section{Motivation}
\label{sec:motivation}

Synchronous compaction has two costs that limit its effectiveness in practice. It can silently degrade accuracy when the compactor drops or distorts information the agent later needs, and it adds substantial latency by blocking task execution while the compactor runs. We examine each in turn.

\newcommand{\dashedlegend}{\tikz[baseline=-0.5ex]{\node[draw, dashed, rounded corners=1pt, inner sep=1.5pt, font=\footnotesize] {Dashed};}}

\begin{figure}[t!]
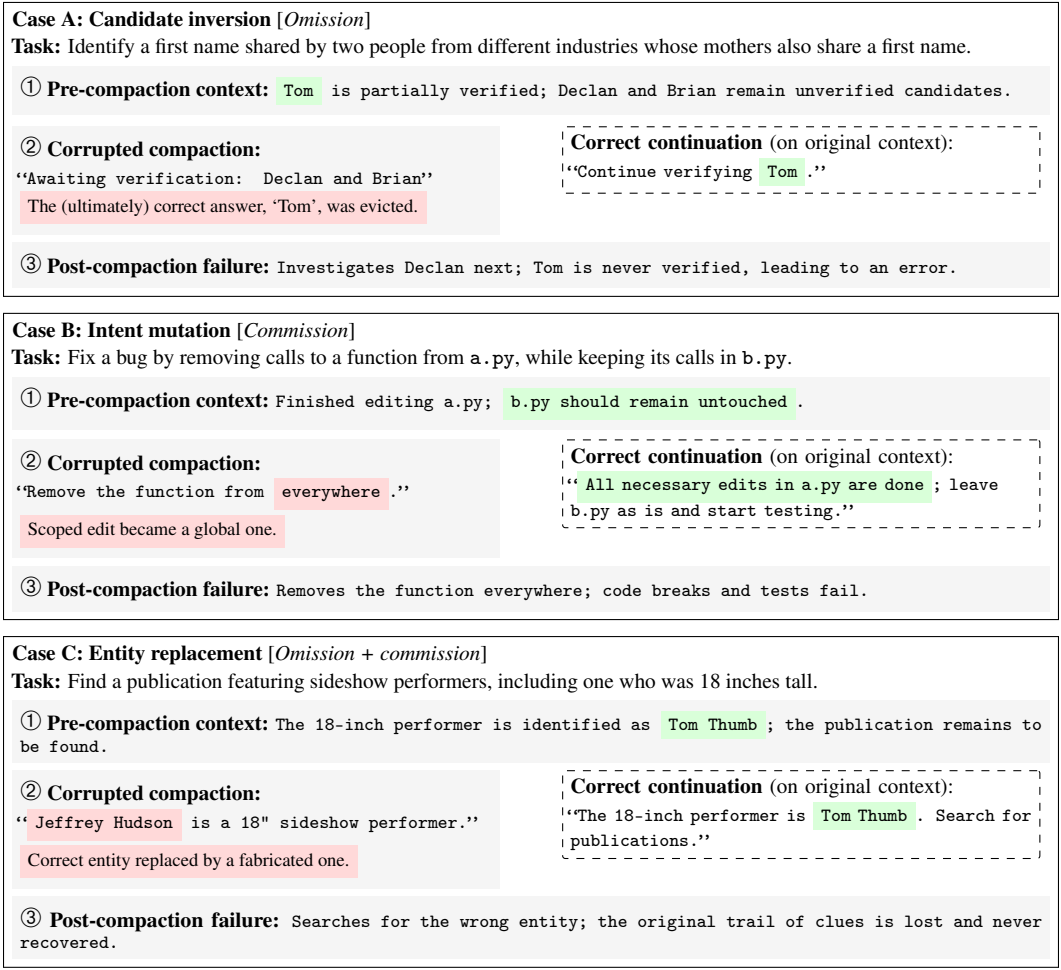

\footnotesize\centering
\setlength{\fboxsep}{3pt}\setlength{\fboxrule}{0.4pt}

\newcommand{\errcase}[7]{%
\noindent\fbox{\begin{minipage}{\dimexpr\linewidth-2\fboxsep-2\fboxrule\relax}
\textbf{#1} [\textit{#2}]\\[1pt]
\textbf{Task:} #3\\[2pt]
\colorbox{gray!8}{\begin{minipage}{\dimexpr\linewidth-2\fboxsep\relax}\vspace{1pt}%
\scalebox{1.3}{\ding{192}}~\textbf{Pre-compaction context:} #4\vspace{1pt}\end{minipage}}\\[2pt]
\noindent\begin{minipage}[t]{0.47\linewidth}
\vspace{0pt}%
\colorbox{gray!8}{\begin{minipage}[t]{\dimexpr\linewidth-2\fboxsep\relax}\vspace{1pt}%
\scalebox{1.3}{\ding{193}}~\textbf{Corrupted compaction:}\\[1pt]
#5\vspace{1pt}\end{minipage}}
\end{minipage}%
\hfill%
\begin{minipage}[t]{0.47\linewidth}
\vspace{0pt}%
\tikz[baseline=(X.north)]{\node[draw, dashed, rounded corners=1pt, inner sep=3pt, text width=\dimexpr\linewidth-10pt\relax] (X) {%
\textbf{Correct continuation} (on original context):\\[1pt]
#6};}%
\end{minipage}\\[2pt]
\colorbox{gray!8}{\begin{minipage}{\dimexpr\linewidth-2\fboxsep\relax}\vspace{1pt}%
\scalebox{1.3}{\ding{194}}~\textbf{Post-compaction failure:} #7\vspace{1pt}\end{minipage}}
\end{minipage}}\vspace{4pt}
}

\errcase{Case~A: Candidate inversion}{Omission}{%
Identify a first name shared by two people from different industries whose mothers also share a first name.}{%
\scriptsize\ttfamily \colorbox{green!15}{Tom} is partially verified; Declan and Brian remain unverified candidates.}{%
{\scriptsize\ttfamily ``Awaiting verification: Declan and Brian''}\\[1pt]
\colorbox{red!15}{\scriptsize The (ultimately) correct answer, `Tom', was evicted.}}{%
{\scriptsize\ttfamily ``Continue verifying \colorbox{green!15}{Tom}.''}}{%
\scriptsize\ttfamily Investigates Declan next; Tom is never verified, leading to an error.}

\errcase{Case~B: Intent mutation}{Commission}{%
Fix a bug by removing calls to a function from \texttt{a.py}, while keeping its calls in \texttt{b.py}.}{%
\scriptsize\ttfamily Finished editing a.py; \colorbox{green!15}{b.py should remain untouched}.}{%
{\scriptsize\ttfamily ``Remove the function from \colorbox{red!15}{everywhere}.''}\\[1pt]
\colorbox{red!15}{\scriptsize Scoped edit became a global one.}}{%
{\scriptsize\ttfamily ``\colorbox{green!15}{All necessary edits in a.py are done}; leave b.py as is and start testing.''}}{%
\scriptsize\ttfamily Removes the function everywhere; code breaks and tests fail.}

\errcase{Case~C: Entity replacement}{Omission + commission}{%
Find a publication featuring sideshow performers, including one who was 18 inches tall.}{%
\scriptsize\ttfamily The 18-inch performer is identified as \colorbox{green!15}{Tom Thumb}; the publication remains to be found.}{%
{\scriptsize\ttfamily ``\colorbox{red!15}{Jeffrey Hudson} is a 18" sideshow performer.''}\\[1pt]
\colorbox{red!15}{\scriptsize Correct entity replaced by a fabricated one.}}{%
{\scriptsize\ttfamily ``The 18-inch performer is \colorbox{green!15}{Tom Thumb}. Search for publications.''}}{%
\scriptsize\ttfamily Searches for the wrong entity; the original trail of clues is lost and never recovered.}
\vspace{-5pt}
\captionsetup{font=small}
\caption{Three illustrative compaction failures from real agent traces on browsing and coding workloads. Each case shows \ding{192} the critical context \textit{before} compaction, \ding{193} the corrupted compacted state, and \ding{194} the resulting failure behavior.
\protect\dashedlegend\,=\, correct continuation (counterfactual; not produced under synchronous compaction);
\colorbox{green!15}{green}\,=\,correct/preserved information; \colorbox{red!15}{red}\,=\,corrupted or injected error.}
\label{fig:error_cases}
\vspace{-15pt}
\end{figure}

\textbf{Silent deviation with unpredictable accuracy impact.} Compaction is lossy by design -- it must decide what to preserve, what to compress, and what to drop. But at compaction time, there is no way to know what information the agent will need later; indeed, discovering the relevant future trajectory is precisely what we employ the agent to autonomously determine. Compaction is therefore inherently a prediction about future relevance made without access to future behavior, and when that prediction is wrong, the summary may omit facts the agent later needs, collapse important intermediate progress into a vague statement, or introduce information the original context did not support.

Figure~\ref{fig:error_cases} provides real examples of such failures from popular coding and web-browsing agents. With omission errors, the summary drops information the agent later needs, e.g., in Case A, the summary silently drops the correct candidate from a set of three, and the agent never recovers it, producing a wrong final answer. With commission errors, the summary introduces or mutates information not supported by the original context, e.g., in Case B, a targeted patch is distorted into a blanket instruction to remove invocations of a function everywhere, leading the agent to make changes the user did not request. Some failures combine both forms, as in Case C, where the summary drops the confirmed entity (``Tom Thumb'') and replaces it with a fabricated one, sending the agent down the wrong search path that never recovers the missing information. Across our workloads, omission accounts for roughly 90\% of failures, with commission and combined failures making up the remainder.

These are not separate pathologies but manifestations of the same underlying problem: the compactor's prediction about what would matter for the agent was incorrect, and synchronous compaction has no mechanism to validate that prediction before committing to it. Put differently, once the summary replaces the original context, the handoff is \emph{irreversible} -- every subsequent step is conditioned on the summary itself, so the agent's continued behavior cannot serve as an independent check. A corrupted summary produces steps consistent with the corruption rather than with the original trajectory, leaving the system without an immediate error signal. The failure is therefore silent, and the accuracy damage only becomes visible when task outcomes degrade.

Existing metrics for traditional summarization fail to address this gap. Source-grounded metrics such as faithfulness~\citep{kryscinski-etal-2020-evaluating}, coverage~\citep{nenkova-passonneau-2004-evaluating}, QA-based probes~\citep{durmus-etal-2020-feqa}, and LLM-as-judge scoring~\citep{liu-etal-2023-geval} evaluate the summary against the original context, but they treat all source facts as equally important and correlate poorly with human judgments of content importance~\citep{deutsch-etal-2022-re, scialom-etal-2021-questeval}. For agent compaction, this manifests as a fundamental tradeoff: reliably preserving the facts the agent \emph{will} need requires near-complete preservation of the source (defeating the purpose of compaction), while any relaxation risks dropping precisely what the agent later needs. The metric has no principled way to prioritize, because facts matter only in relation to \emph{future agent behavior}, which it cannot see. This is particularly damaging since omissions dominate the error distribution but produce no contradiction with the source for these metrics to detect~\citep{zhang-etal-2023-extractive, kim-etal-2024-fables}. Reference-grounded metrics like ROUGE~\citep{lin-2004-rouge} are inapplicable since compaction has no reference summary.

Taken together, what existing metrics miss is \emph{sufficiency}: a compacted state must preserve enough of the original context for the continuing agent to avoid deviating from the trajectory it would have followed without compaction. Sufficiency depends on future agent behavior, which is precisely what source-grounded validation cannot see. This motivates a new validation signal -- one grounded in the agent behavior that compacted state is meant to support, rather than in the source context it replaces.

\begin{figure}[t]
    \centering
    \begin{subfigure}[b]{0.53\linewidth}
        \centering
        \includegraphics[width=\linewidth]{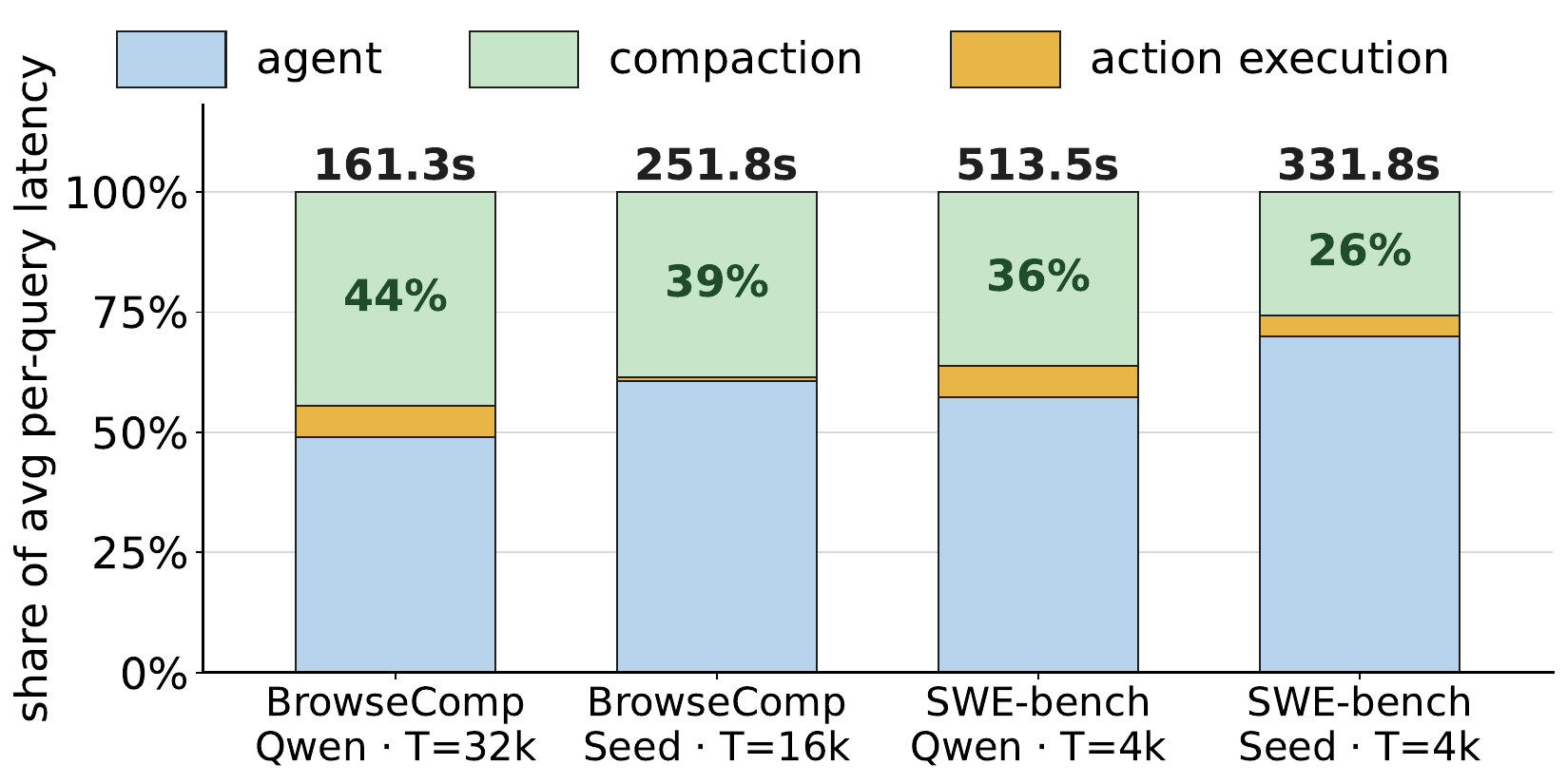}
        \caption{\footnotesize{Normalized latency breakdown of synchronous compaction. Each workload runs with either Qwen3.5-9B or Seed-OSS-36B-Instruct as the agent, with different compaction thresholds (\textbf{T}).}}
        \label{fig:compaction_ratio}
    \end{subfigure}
    \hfill
    \begin{subfigure}[b]{0.43\linewidth}
        \centering
        \includegraphics[width=\linewidth]{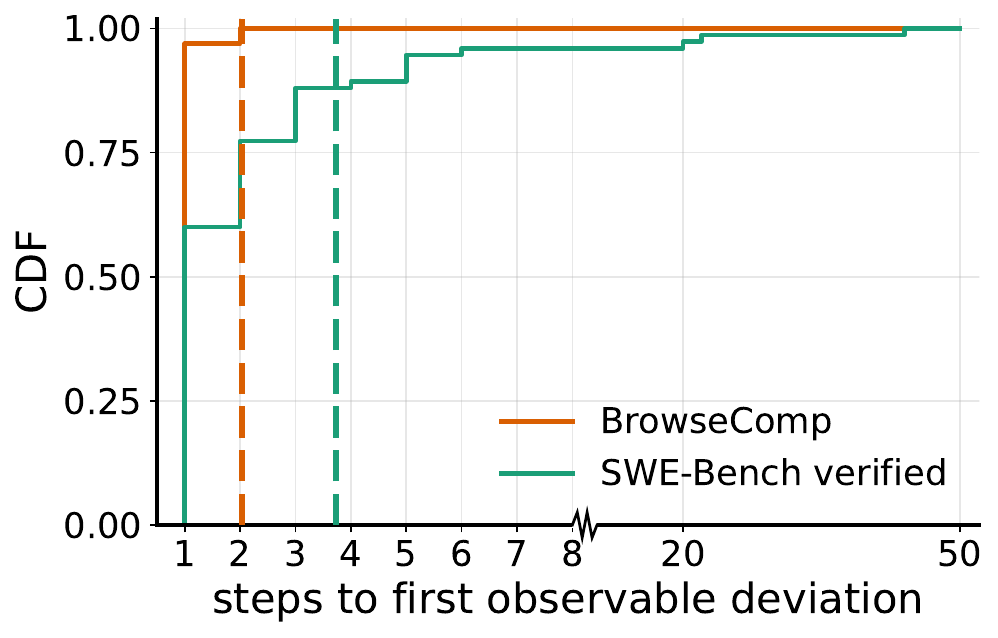}
        \caption{\footnotesize{Locality of compaction-induced silent deviation.}}
        \label{fig:compaction_induced_silent_deviation}
    \end{subfigure}
    \captionsetup{font=small}
    \caption{Challenges with synchronous compaction. (a)~Compaction lies on the agent's critical path, accounting for 26--44\% of end-to-end latency despite producing no task-useful action. (b) Distribution of agent steps between compaction and the first observable trajectory deviation due to compaction errors; most deviations appear within the first few steps, well within Slipstream's typical async window (dashed lines).}
    \label{fig:motivation_measurements}
    \vspace{-20pt}
\end{figure}

\textbf{Blocking execution and high latencies.} Beyond its accuracy risk, synchronous compaction stalls task progress. While the compactor LLM runs, the agent cannot issue tool calls, perform reasoning, or make progress on the task. Each compaction event requires an LLM inference call but produces no task-useful actions. Moreover, this overhead compounds over long-horizon tasks, especially since modern systems trigger compaction aggressively to stay ahead of context rot rather than waiting until the context window fills. Figure~\ref{fig:compaction_ratio} breaks down the (normalized) end-to-end latencies
when using synchronous compaction at varying frequencies across our agent workloads. As shown, compaction accounts for an average of 26-44~\% of total per-query latency across the considered settings. 
\vspace{-15pt}

\section{Method}
\label{sec:method}
\vspace{-5pt}

\system{} turns compaction into an asynchronous validation procedure: when compaction is triggered, the agent continues executing on the original, uncompacted context while a compactor runs in parallel on a separate thread. Because both processes operate from the same pre-compaction input, the candidate summary and the agent's continued behavior are generated independently, providing the held-out validation signal missing under synchronous execution (Section~\ref{sec:motivation}). \system{} uses the same context-length trigger as the synchronous baseline, so any performance difference comes from how compaction is scheduled and validated rather than when it occurs. We refer to the agent's continued execution during compaction (i.e., its tool calls and reasoning steps) as the \emph{next-$k$ trajectory}. Once the compactor returns, a trajectory-grounded judge analyzes the summary against the next-$k$ trajectory and either accepts the candidate for adoption or returns a diagnosis used to update it before adoption. Figure~\ref{fig:overview} illustrates this execution model and contrasts it with the synchronous baseline.
\vspace{-10pt}

\subsection{Why a finite trajectory window suffices}
\label{sec:method-locality}

\system{} observes only a finite continuation of the agent's execution -- the next-$k$ steps generated while compaction runs. The effective $k$ is not a fixed parameter but is instead naturally determined by each compaction duration relative to agent step time. Compaction is reliably longer than a single agent step because it generates a substantial summary, while agent steps typically generate only short reasoning traces and tool calls. In our evaluations, compaction overlaps with an average of 2.1 agent steps on BrowseComp and 3.7 steps on SWE-bench Verified.

To assess whether this window is sufficient for validation, we measure how quickly compaction errors manifest in agent behavior (Figure~\ref{fig:compaction_induced_silent_deviation}). On BrowseComp, where search-style tasks tightly couple context to action, 97\% of first error manifestations occur at $k=1$ and 100\% by $k=2$, well within \system{}'s average window. In contrast, for SWE-bench Verified where coding tasks involve intermediate reasoning chains before actions commit to corrupted information, the distribution is broader: 60\% of errors at $k=1$, 88\% by $k=3$, and 96\% by $k=6$. Here, windows with \system{} cover 88\% of first error manifestations; the few errors that surface outside of \system{}'s validation window (and thus, only after the compaction summary is applied) cannot be recovered, as with synchronous compaction. Overall, both distributions exhibit a \emph{locality property}: the first observable error manifests soon after compaction, within the agent's current subgoal, allowing the (finite) asynchronous window to capture them.
\vspace{-10pt}

\subsection{Validating summaries with the trajectory-grounded judge}
\label{sec:method-judge}

\system{}'s judge decides whether the candidate compacted state is acceptable, decomposing this into two complementary checks against the next-$k$ trajectory: one verifies that the compacted state is consistent with the agent's downstream intentions, and the other verifies that it preserves the specific information those steps rely on. The reason is that these two properties can fail independently -- a summary can preserve every relevant fact while changing the agent's forward intent, or preserve intent while losing the facts needed to act on it. We describe each check below:
\squishlist
\item The \emph{statement-level check} asks whether the compacted state preserves the concrete facts, constraints, intermediate results, and tool observations explicitly used in the next-$k$ trajectory. The agent's thinking tokens reveal not only the facts driving its immediate actions, but also facts it previously relied on along the current trajectory. This check targets content errors exemplified in Figure~\ref{fig:error_cases}, such as the omissions in Case~A and the entity-level corruptions in Case~C.
\item The \emph{plan-level check} asks whether the compacted state supports the same forward intent as the original continuation. This targets structural drift: the summary may preserve isolated facts while steering the agent toward a different next action, as in Case~B. The check is feasible because long-horizon agents articulate their plans explicitly to stay coherent across many steps. They typically do this in one of two ways: through coarse end-to-end framing at task start that is periodically refined with shorter-term plans as the agent acts (``I will first locate the affected imports, then modify each file'') as in SWE-Bench and GAIA~\citep{gaia}, or through an explicit (global) running checklist that updates with progress as in BrowseComp. The next-$k$ trajectory reveals the current plan, enabling direct validation on articulated intents. If plans are revised within the next-$k$ window, the judge validates against the most recent articulation.

\squishend

On accept, the judge passes the candidate to the adoption procedure (\S~\ref{sec:method-adoption}); on reject, it returns a brief diagnosis of the missing or corrupted trajectory-relevant information. By default, the judge uses the same model as the agent with a structured prompt (Appendix~\ref{sec:judge-prompt}); Section~\ref{sec:experiments} shows that judge validation is largely insensitive to model choice and contributes materially beyond asynchrony alone.

\vspace{-5pt}

\subsection{Adoption and targeted update}
\label{sec:method-adoption}

On judge acceptance, \system{} swaps execution to the compacted state augmented with the next-$k$ continuation steps that completed during compaction. Because the judge has already verified consistency, the live agent can safely continue from the same task progress with a reduced context. On rejection, \system{} briefly pauses agent execution and performs a targeted update to the compacted state, using the judge's diagnosis and the relevant evidence from the uncompacted continuation to repair the specific trajectory-relevant omission or corruption -- e.g., restoring a dropped constraint, correcting a mutated entity, or preserving the latest patch intent -- without running synchronous compaction from scratch. \system{} then adopts the updated compacted state together with the next-$k$ continuation steps. Empirically, rejection is rare (1.0--3.5\% on BrowseComp, 5.4--8.5\% on SWE-bench Verified), so \system{} retains most of the latency benefit while using targeted updates as a guardrail against summaries the judge identified as trajectory-changing. In the rare cases where targeted update cannot recover (e.g., structural drift that is too pervasive to repair locally), \system{} falls back to synchronous compaction.
\vspace{-10pt}

\subsection{System considerations}
\label{sec:method-system}

\system{} exploits structural properties of long-horizon agent serving to make asynchronous execution practical. The key enabler is that modern long-horizon agent serving is constrained by memory rather than compute~\citep{memory_walls, vllm}: effective batch sizes sit well below the available compute limit, leaving headroom to run the compactor and judge alongside the main agent at minimal additional cost. Within this headroom, \system{} further optimizes its use of memory and compute. The agent thread and the compaction thread operate over the same pre-compaction context, so \system{} reuses the shared prefix in the KV cache instead of materializing two independent copies, similar to prior work on shared-prefix inference~\citep{specreason}. To avoid duplicating memory loads over this prefix, we adopt multi-layer cascade attention~\citep{flashinfer_cascade_inference}, which loads the prefix once and lets both threads attend to it without redundant memory traffic. The judge's inputs are also lightweight: it processes only the candidate compacted state and the next-$k$ trajectory (typically 3.6-6.8\% of the pre-compaction context size) and outputs a brief accept/reject decision with a diagnosis on rejection. %
\vspace{-10pt}

\section{Experiments}
\label{sec:experiments}
\vspace{-5pt}
\subsection{Setup}

\begin{table}[t]
    \centering
    \captionsetup{font=footnotesize}
    \caption{\textbf{\system{}}'s (absolute) accuracy improvement over the Sync baseline across models, workloads, and compaction thresholds. Thresh. denotes the active-context token count at which compaction is triggered. \system{} improves accuracy in every configuration (by up to 8.8\%).}
    \vspace{0.5em}
    \label{tab:accuracy}
    \newcommand{\accgain}[1]{\,\scriptsize{(#1)}}
    \begin{tabular}{ll@{\hskip 0.8em}lcc@{\hskip 1.2em}lcc}
        \toprule
        & & \multicolumn{3}{c}{\textbf{SWE-bench verified}} & \multicolumn{3}{c}{\textbf{BrowseComp}} \\
        \cmidrule(lr){3-5} \cmidrule(l){6-8}
        \textbf{Model} & & \textbf{Thresh.} & \textbf{Sync} & \textbf{\system{}} & \textbf{Thresh.} & \textbf{Sync} & \textbf{\system{}} \\
        \midrule
        \multirow{3}{*}{Qwen3.5-9B}
            & & 4k & $23.4\%$ & $\mathbf{29.8\%}$ \accgain{+$6.4\%$} & 32k & $53.3\%$ & $\mathbf{57.3\%}$ \accgain{+$4.0\%$} \\
            & & 6k & $22.0\%$ & $\mathbf{30.8\%}$ \accgain{+$8.8\%$} & 48k & $52.7\%$ & $\mathbf{57.3\%}$ \accgain{+$4.6\%$} \\
            & & 8k & $20.4\%$ & $\mathbf{29.2\%}$ \accgain{+$8.8\%$} & 64k & $55.3\%$ & $\mathbf{58.0\%}$ \accgain{+$2.7\%$} \\
        \midrule
        \multirow{3}{*}{Seed-OSS-36B}
            & & 4k & $29.6\%$ & $\mathbf{35.8\%}$ \accgain{+$6.2\%$} & 16k & $46.7\%$ & $\mathbf{48.7\%}$ \accgain{+$2.0\%$} \\
            & & 6k & $30.6\%$ & $\mathbf{37.4\%}$ \accgain{+$6.6\%$} & 24k & $52.0\%$ & $\mathbf{53.3\%}$ \accgain{+$1.3\%$} \\
            & & 8k & $38.0\%$ & $\mathbf{40.6\%}$ \accgain{+$2.6\%$} & 32k & $49.3\%$ & $\mathbf{52.7\%}$ \accgain{+$3.4\%$} \\
        \bottomrule
    \end{tabular}
    \vspace{-10pt}
\end{table}

\textbf{Models and workloads.} We evaluate \system{} using two models instruction-tuned for long-horizon agentic tasks: \textbf{Qwen3.5-9B}~\citep{qwen3}, a hybrid MoE model, and \textbf{Seed-OSS-36B-Instruct}~\citep{seed_oss}, a dense Transformer.
We evaluate on two representative long-horizon agent workloads. \textbf{BrowseComp}~\citep{browsecomp} evaluates web-browsing agents that must maintain search progress and entity constraints across many tool calls. \textbf{SWE-bench verified}~\citep{swebench} evaluates coding agents that must preserve issue context, test outcomes, and repository observations over multi-step debugging trajectories. These workloads stress different forms of trajectory state: BrowseComp primarily requires search-state maintenance, while SWE-bench requires preserving both factual observations and evolving code-editing plans.

\textbf{Baselines, methodology, and hardware.}
We compare \system{} with a synchronous compaction baseline (\textbf{Sync}) using the same compaction thresholds. In both schemes, compaction is invoked when the active context reaches the threshold. The difference is that Sync blocks agent execution and always adopts the compacted context, whereas \system{} performs compaction asynchronously and validates or updates the candidate compacted state using the next $k$ trajectory steps (Figure~\ref{fig:overview}). Unless otherwise stated, all schemes use the same agent prompt, tool interface, and compactor prompt for a given model and workload. For each model-workload pair, we evaluate three compaction thresholds: a default threshold set to approximately $\sim$20\% of the average full trajectory length, plus thresholds at $\frac{2}{3}\times$ and $\frac{4}{3}\times$ the default to assess sensitivity to the compaction frequency.
Across all evaluations, the trajectory-grounded judge accepts a compacted state if its score is at least 7 out of 10; candidates below this threshold trigger the targeted update path (Section~\ref{sec:method-adoption}). We report task success rate as accuracy, and report average per-query latency decomposed into agent reasoning, action execution, compaction, and judge/update overhead.
We serve all models locally using vLLM 0.19.1 in our experiments. \texttt{Qwen3.5-9B} runs on one NVIDIA H100 80GB GPU, while \texttt{Seed-OSS-36B-Instruct} runs on two.
\vspace{-5pt}

\subsection{Main Results} \label{sec:main-results}

\begin{figure}[!t]
    \centering
    \begin{subfigure}[b]{\linewidth}
        \centering
        \includegraphics[width=\linewidth]{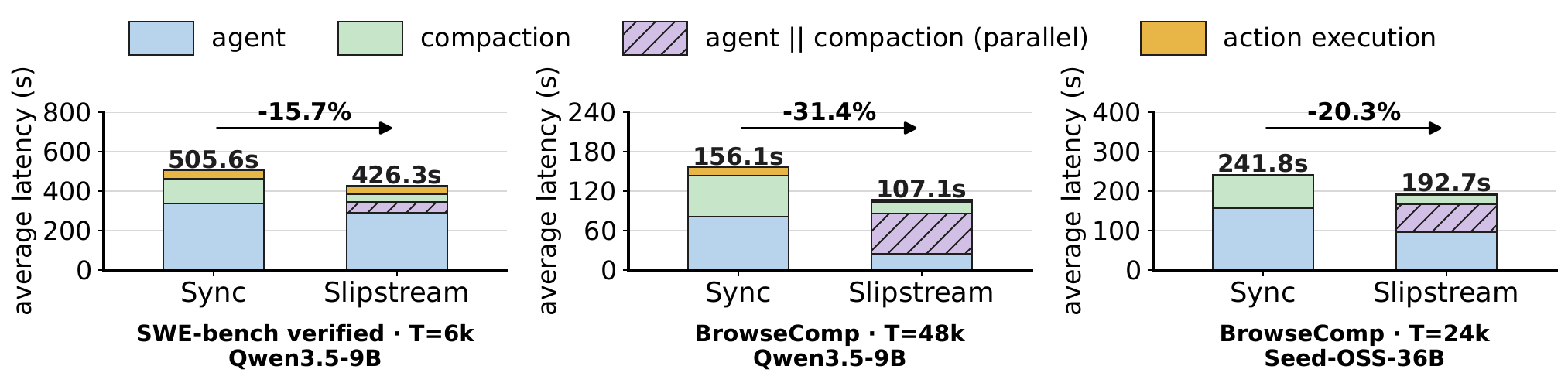}
        \caption{\footnotesize{Per-query latency breakdown for three representative configurations, decomposing total runtime into agent reasoning, action execution, synchronous compaction (Sync only), and overlapped agent–compaction execution (\system{} only). We omit judge/update overhead as it is negligible ($<1\%$ of total latency across evaluations).}}
        \label{fig:latency_breakdown}
    \end{subfigure}
    \begin{subfigure}[b]{\linewidth}
        \centering
        \includegraphics[width=\linewidth]{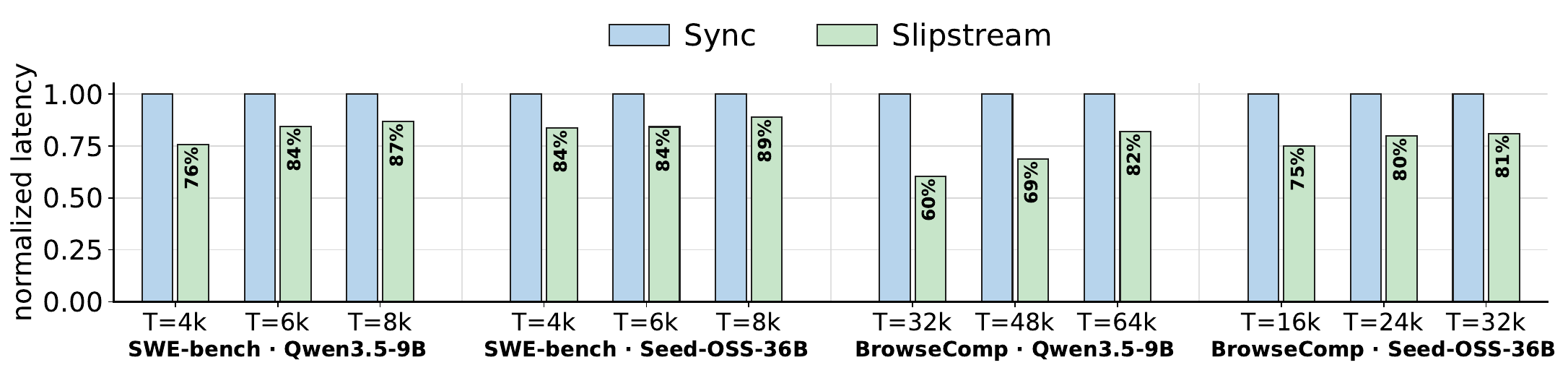}
        \caption{\footnotesize{Normalized end-to-end latency (Slipstream / Sync) across all evaluated configurations.}}
        \label{fig:latency_compact}
        \vspace{-6pt}
    \end{subfigure}
    \captionsetup{font=small}
    \caption{Latency analysis across workloads, models, and compaction thresholds (\textbf{T}). \system{} consistently reduces latency by 11.3-39.7\% via moving compaction off the critical path, with greater reductions at smaller thresholds, where compaction is invoked more frequently. Full results are in Appendix~\ref{sec:detailed-breakdown}.}
    \label{fig:latency_analysis}
    \vspace{-15pt}
\end{figure}

\textbf{Accuracy.} Table~\ref{tab:accuracy} shows that \system{} improves task success rate over synchronous compaction across all model-workload-threshold combinations.
On BrowseComp, \system{} improves accuracy by 1.3--4.6\%, suggesting that it better preserves active search state and avoids silent repetitions or unintended changes in search direction after compaction.
On SWE-bench verified, the gains are larger (2.6--8.8\%), indicating that trajectory-grounded validation is especially useful for preserving patch intent, test interpretation, and repository observations that synchronous compaction may distort or omit.

\textbf{Latency.} Figure~\ref{fig:latency_breakdown} shows detailed latency breakdowns for three representative workload--model combinations at their default compaction thresholds (breakdown results for all settings are in Appendix~\ref{sec:detailed-breakdown}). In the Sync baseline, compaction appears as a blocking component on the critical path. In contrast, \system{} overlaps compaction with continued agent execution on the original context, so most of the compaction time is hidden within the asynchronous execution window and contributes little net latency overhead. Figure~\ref{fig:latency_compact} summarizes the end-to-end effect across all evaluated settings: \system{} consistently reduces average per-query latency relative to Sync.
\system{}'s relative latency reduction grows as compaction is invoked more frequently, since blocking summarization then accounts for a larger share of end-to-end runtime.

Overall, the results show that asynchrony is not merely an efficiency optimization. By continuing on the original context while compaction runs, \system{} obtains a held-out trajectory signal that can be used to validate and improve the compacted state before adoption, yielding consistent accuracy gains while also hiding summarization latency.
The absence of a monotonic accuracy trend across thresholds reflects a known tension: more frequent compaction better mitigates context rot but also creates more opportunities for silent deviation (Section~\ref{sec:motivation}).
Slipstream's per-compaction validation makes it robust to this tradeoff.

\subsection{Microbenchmarks}

\begin{wrapfigure}{r}{0.5\textwidth}
    \vspace{-5pt}
    \centering
    \includegraphics[width=\linewidth]{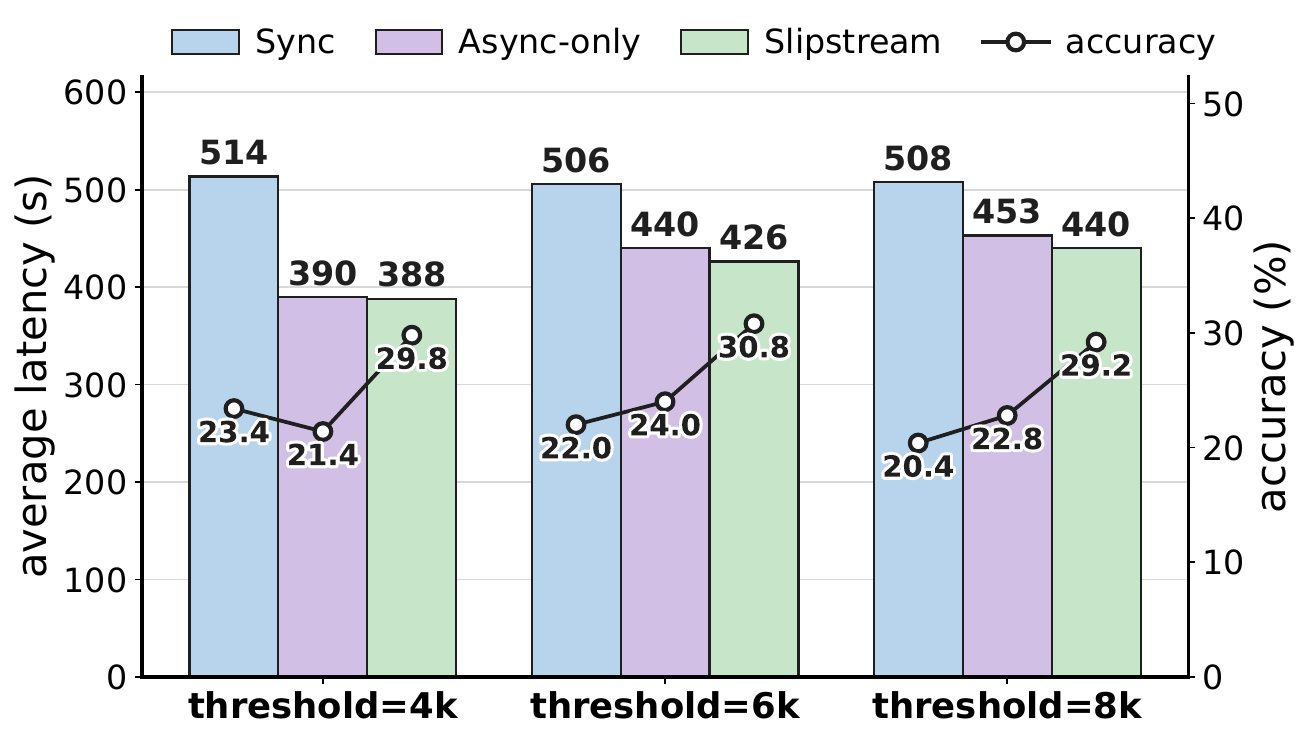}
    \captionsetup{font=small}
    \caption{[SWE-bench verified, Qwen3.5-9B] Isolating the contributions of asynchrony and judging. Async-only captures most of Slipstream's latency benefit but does not improve accuracy over Sync, while full Slipstream retains the latency benefit and improves accuracy across all thresholds -- indicating that the accuracy gain comes from trajectory-grounded validation rather than asynchrony alone.}
    \label{fig:validation_ablation_swe_qwen}
    \vspace{-5pt}
\end{wrapfigure}

\textbf{Ablation studies on accuracy.}
Compared to Sync, \system{} introduces two changes with potential accuracy implications: it generates the next-k agent actions on the uncompacted context asynchronously, and it validates and optionally repairs the candidate summary before adoption. To isolate which of these drives the accuracy gains, we compare Sync and \system{} against a third scheme, Async-only, which overlaps compaction with agent execution on the uncompacted context but always accepts the resulting summary without validation. As shown in Figure~\ref{fig:validation_ablation_swe_qwen}, Async-only captures most of \system{}'s latency benefit over Sync, but its accuracy remains close to Sync across all three thresholds -- overlapping execution alone does not recover the accuracy lost to corrupted summaries, since Async-only still adopts every compacted state regardless of whether it preserves trajectory-relevant information. \system{} achieves the greatest latency reduction while improving accuracy by 6.4-8.8\% over Sync, showing that the accuracy gain comes from trajectory-grounded validation rather than from asynchrony alone.

\begin{wraptable}{r}{0.50\textwidth}
    \centering
    \vspace{-1em}
    \newcommand{\accgain}[1]{\,\scriptsize{(#1)}}
    \begin{tabular}{lll}
        \toprule
        \textbf{Threshold} & \textbf{Judge} & \textbf{Accuracy} \\
        \midrule
        \multirow{4}{*}{4k} & Sync (no judge) & $23.4\%$ \\
                             \cmidrule(l){2-3}
                             & Qwen3.5-9B & $29.8\%$ \accgain{+$6.4\%$} \\
                             & Qwen3.5-2B & $28.2\%$ \accgain{+$4.8\%$} \\
                             & Llama-3.2-3B & $28.2\%$ \accgain{+$4.8\%$} \\
        \midrule
        \multirow{4}{*}{6k} & Sync (no judge) & $22.0\%$ \\
                             \cmidrule(l){2-3}
                             & Qwen3.5-9B & $30.8\%$ \accgain{+$8.8\%$} \\
                             & Qwen3.5-2B & $28.6\%$ \accgain{+$6.6\%$} \\
                             & Llama-3.2-3B & $31.8\%$ \accgain{+$9.8\%$} \\
        \midrule
        \multirow{4}{*}{8k} & Sync (no judge) & $20.4\%$ \\
                             \cmidrule(l){2-3}
                             & Qwen3.5-9B & $29.2\%$ \accgain{+$8.8\%$} \\
                             & Qwen3.5-2B & $28.0\%$ \accgain{+$7.6\%$} \\
                             & Llama-3.2-3B & $27.2\%$ \accgain{+$6.8\%$} \\
        \bottomrule
    \end{tabular}
    \captionsetup{font=small}
    \caption{[SWE-bench verified] Varying the judge model used in \system{} with a fixed Qwen3.5-9B agent. Smaller and cross-family judges consistently outperform Sync and largely retain the gains of the same-model judge, indicating trajectory-grounded validation does not require a judge as capable as the agent.}
    \label{tab:judge_ablation_swe_qwen}
    \vspace{-10pt}
\end{wraptable}

\textbf{Sensitivity to judge model choice.}
By default, \system{} uses the same model for the agent and the judge. This is convenient at the system level -- both threads share a prefix and a model instance, avoiding redundant prefill during judging and eliminating the need to host a second model (Section~\ref{sec:method-system}). But it leaves open whether judging actually requires a model as capable as the agent, or whether a smaller, cheaper judge would suffice. To study this, we keep the agent model fixed at Qwen3.5-9B on SWE-bench verified and vary only the judge, swapping in two smaller models from different families.
Table~\ref{tab:judge_ablation_swe_qwen} shows that these weaker judges consistently outperform synchronous compaction and largely retain the accuracy gains of the same-model judge. This suggests that the gains come from the trajectory-grounded validation signal itself, rather than from relying on an especially strong judge model. It also broadens practical deployment options: the judge can be served by a cheaper model, placed in a separate engine, or scaled independently from the main agent.
\vspace{-5pt}

\section{Conclusion}
\label{sec:conclusion}
\vspace{-5pt}
We present \system{}, a trajectory-grounded compaction system for long-horizon LLM agents. Compaction faithfulness has no synchronous ground truth: the compactor must predict what the agent will need next, and synchronous execution offers no way to check that prediction before committing to it. Asynchrony resolves this. Running the compactor in parallel with continued agent execution generates the summary and the agent's behavior independently from the same input, producing a validation signal no synchronous configuration can construct. \system{} improves task accuracy by up to 8.8 percentage points over synchronous compaction while cutting end-to-end latency by up to 39.7\%. More broadly, running auxiliary procedures synchronously on an agent's critical path can foreclose validation signals that asynchrony makes available -- a pattern worth examining beyond compaction itself.

\begin{ack}
We thank Princeton’s Systems for Artificial Intelligence Lab (SAIL) and Princeton Language and
Intelligence (PLI) for providing the hardware resources for running experiments. This work was
supported by NSF CNS grants 2147909, 2151630, 2140552, 2153449, and 2152313.
\end{ack}

\bibliographystyle{plainnat}
\bibliography{references}

\newpage

\appendix

\section{Detailed latency breakdown for all combinations in \S~\ref{sec:main-results}}
\label{sec:detailed-breakdown}

\begin{figure}[!htbp]
    \centering
    \includegraphics[width=0.9\linewidth, keepaspectratio]{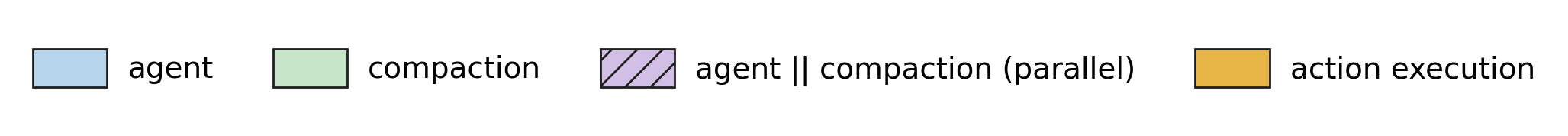}
    \vspace{0.2em}
    \begin{subfigure}[b]{\linewidth}
        \includegraphics[width=\linewidth, height=0.2\textheight, keepaspectratio]{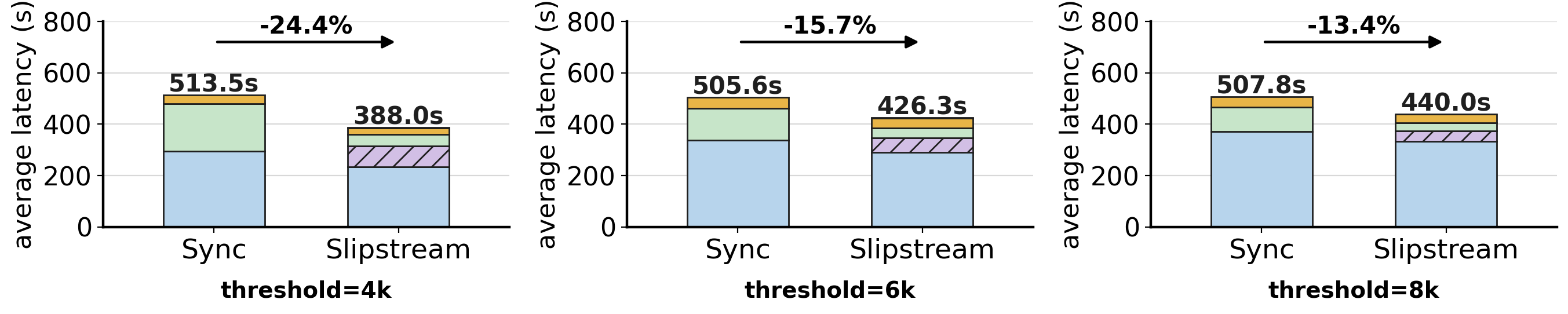}
        \caption{SWE-bench verified, Qwen3.5-9B}
        \label{fig:latency_SWE_Qwen}
    \end{subfigure}
    \begin{subfigure}[b]{\linewidth}
        \includegraphics[width=\linewidth, height=0.2\textheight, keepaspectratio]{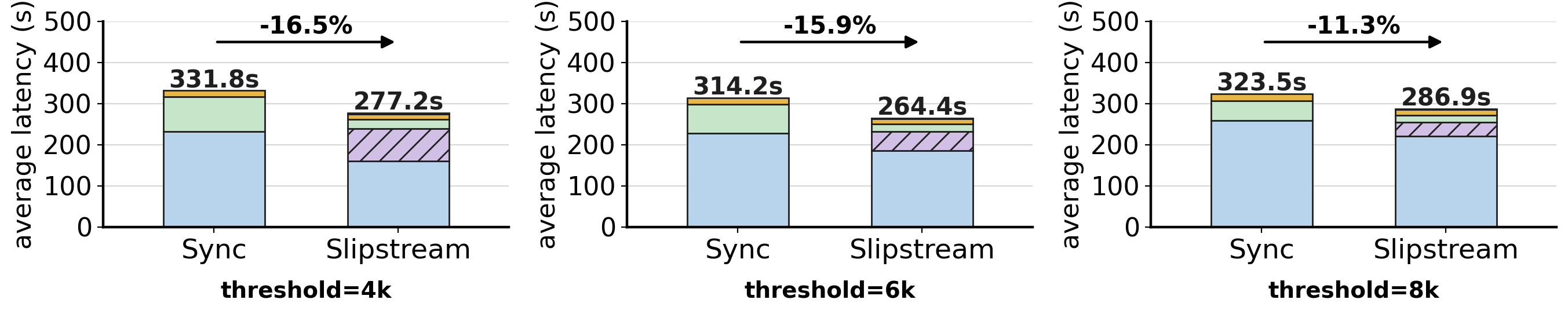}
        \caption{SWE-bench verified, Seed-OSS-36B-Instruct}
        \label{fig:latency_SWE_Seed}
    \end{subfigure}
    \begin{subfigure}[b]{\linewidth}
        \includegraphics[width=\linewidth, height=0.2\textheight, keepaspectratio]{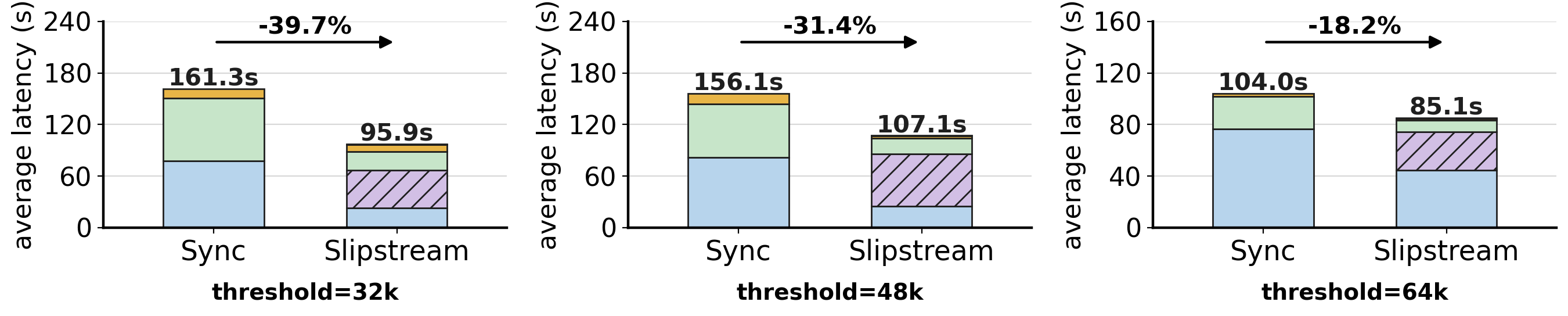}
        \caption{BrowseComp, Qwen3.5-9B}
        \label{fig:latency_BC_Qwen}
    \end{subfigure}
    \begin{subfigure}[b]{\linewidth}
        \includegraphics[width=\linewidth, height=0.2\textheight, keepaspectratio]{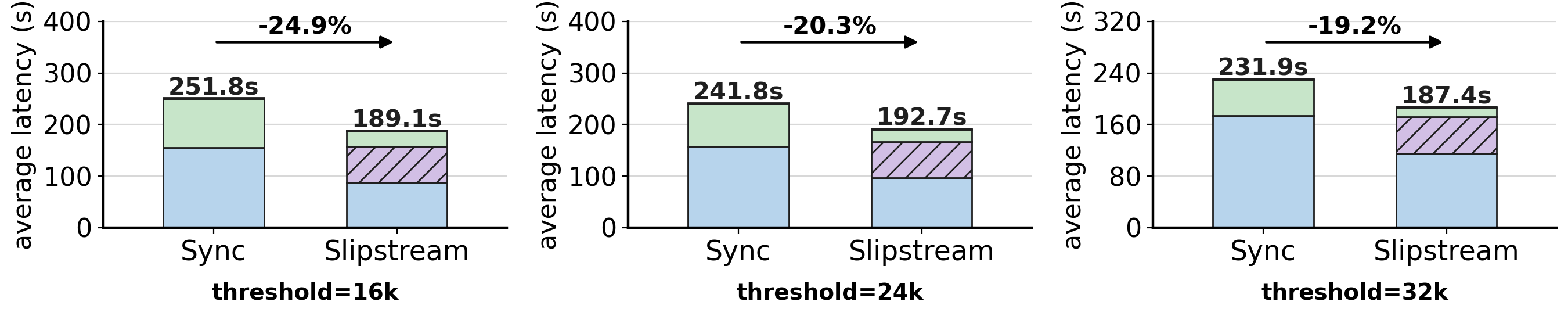}
        \caption{BrowseComp, Seed-OSS-36B-Instruct}
        \label{fig:latency_BC_Seed}
    \end{subfigure}
    \caption{\small{Latency breakdown across workloads, models, and compaction thresholds. \system{} hides summarization overhead within the asynchronous execution window, achieving near-zero net latency overhead.}}
    \label{fig:latency_breakdown_full}
    \vspace{-15pt}
\end{figure}

\section{Detailed Prompts}
\label{sec:prompt-engineering}

\subsection{Trajectory-Grounded Judge Prompt}
\label{sec:judge-prompt}

The judge decides whether a candidate compacted state is sufficient to adopt by checking it against the next-$k$ trajectory---the agent's continued execution on the uncompacted context while compaction ran asynchronously (\S~\ref{sec:method-judge}). It decomposes this into two complementary checks: a \emph{plan-level} check (whether the compacted state supports the same forward intent as the continuation) and a \emph{statement-level} check (whether the compacted state preserves the concrete facts, constraints, and observations the continuation relies on). The prompt returns a structured JSON verdict; on rejection, the diagnosis feeds into the targeted update path (\S~\ref{sec:method-adoption}).

\begin{lstlisting}
You are evaluating whether a candidate compacted summary is consistent
with the agent's speculative steps -- the actions the agent took on
the original, uncompacted context while compaction ran asynchronously.

The agent's new context after adoption will be:
  [candidate compacted summary] + [speculative trajectory appended verbatim]

Judge consistency on TWO criteria using ONLY the candidate compacted
summary and the speculative trajectory provided below.

---
CANDIDATE COMPACTED STATE:
{summary_handover}

---
SPECULATIVE TRAJECTORY (reasoning + tool calls taken during async compaction):
{spec_actions}

---
CRITERIA:

1. PLAN ALIGNMENT
   The compacted state contains open/pending items (marked [PENDING],
   [OPEN], or described as unresolved/unconfirmed). Does the next-k
   trajectory pursue these open items with the same forward intent?
   - 10: Trajectory directly targets the most critical open item(s)
   - 7-9: Trajectory targets an open item, but not the most critical
   - 4-6: Trajectory is topically related but takes an indirect path
   - 1-3: Trajectory addresses something not listed as open/pending
   - 0:   Trajectory is unrelated to any compacted-state content,
          OR compacted state has no open items but trajectory
          continues searching instead of finishing

2. INFORMATION PRESERVATION
   The speculative trajectory's reasoning may reference prior context
   (entities, partial findings, verified facts, tool observations).
   Are those concrete facts preserved in the candidate compacted state?
   - 10: Every entity/fact the trajectory relies on appears in the
         compacted state
   - 7-9: Most referenced information is present; minor details
          missing but reconstructable from context
   - 4-6: Some key information the trajectory depends on is missing,
          which could cause confusion after adoption
   - 1-3: Trajectory relies heavily on information not in the
          compacted state
   - 0:   Trajectory references a completely different task state
          than what the compacted state describes

Respond with JSON only (no markdown, no extra text):
{"plan_alignment": <int 0-10>,
 "information_preservation": <int 0-10>,
 "score": <int 0-10>,
 "reasoning": "<one sentence on the key strength or weakness>"}

The overall score must equal:
  round(0.5 * plan_alignment + 0.5 * information_preservation)
\end{lstlisting}

\subsection{SWE-bench verified Workflow}

Our prompt for SWE-bench follows OpenHands~\citep{openhands}. The full system prompt is shown below.

\begin{lstlisting}
You are a helpful assistant that can interact with a computer to solve
programming tasks.

ROLE
  Your primary role is to assist users by executing commands, modifying
  code, and solving technical problems effectively. You should be
  thorough, methodical, and prioritize quality over speed.
  * If the user asks a question, like "why is X happening", don't try
    to fix the problem. Just give an answer to the question.

FILE SYSTEM GUIDELINES
  * When a user provides a file path, do NOT assume it's relative to
    the current working directory. First explore the file system to
    locate the file before working on it.
  * If asked to edit a file, edit the file directly, rather than
    creating a new file with a different filename.
  * For edits to existing source files, use targeted in-place edits
    via sed or inline python. Do NOT rewrite an existing file via
    heredoc.
  * Only /testbed/reproduce_issue.py is allowed as a new top-level
    script. To fix the bug, edit files inside /testbed/<package>/...
    directly.
  * Do NOT use AST-based source patching. Use sed -i or inline
    python str.replace only.
  * If a prior edit corrupted a source file, restore it with
    git checkout.

TROUBLESHOOTING
  * If repeated attempts fail, do NOT keep adjusting the same
    approach. Step back and explicitly enumerate 5-7 different
    possible root causes. Rank by likelihood and address the most
    likely first.
  * When a test fails after a fix attempt, do not rewrite the test
    or source file from scratch. Instead, (1) inspect intermediate
    values, (2) identify the specific line that produces the wrong
    value, (3) make one targeted edit at that line.

CODE QUALITY
  * Focus on making the minimal changes needed to solve the problem.
  * Place all imports at the top of the file unless this would cause
    issues (e.g., circular imports).

NETWORK
  * The environment has NO network access. NEVER run commands that
    reach the internet (pip install, apt install, git clone, curl,
    wget, etc.).
  * All dependencies needed to run the tests are already installed.
  * If an import fails, diagnose with which python and check the
    environment. Do not attempt to install missing packages.

SUBMISSION
  * You have a fixed step budget; spend it on edits, not
    verification.
  * If git diff shows source-file edits that plausibly address the
    issue, submit.
  * Running additional tests after a plausible fix wastes budget.
  * Only continue editing if git diff is empty or clearly does not
    address the task.

CONTINUATION
  * The conversation history is authoritative -- every command listed
    there has already run and every edit described there is already
    on disk. Do not re-run the same command.
  * Before applying an edit, a short sed -n to confirm the pattern
    still matches is permitted.
\end{lstlisting}

\subsection{BrowseComp Workflow}

Our prompt for BrowseComp is adopted and modified from \citep{context_folding}. It instructs the agent to follow a four-phase research protocol (deconstruction, iterative search, synthesis, verification) and to use the \texttt{search}, \texttt{open\_page}, and \texttt{finish} tools. The full system and user prompts are shown below.

\begin{lstlisting}
You are a meticulous and strategic research agent. Your primary
function is to conduct comprehensive, multi-step research to deliver
a thorough, accurate, and well-supported report in response to the
user's query.

Core principles:
  * Rigor: Execute every step with precision and attention to detail.
  * Objectivity: Synthesize based on evidence, not assumptions.
    Note and investigate conflicting information.
  * Thoroughness: Never settle for a surface-level answer. Always
    strive to uncover underlying details, context, and data.
  * Transparency: Your reasoning should be clear at every step,
    linking evidence directly to conclusions.

Follow this structured protocol to find the answer:

Phase 1: Deconstruction & Strategy
  1.1 Deconstruct the Query:
      - Identify the core question(s).
      - Isolate key entities, concepts, and relationships.
      - List all constraints, conditions, and required data points.
  1.2 Hypothesize & Brainstorm:
      - Brainstorm search vectors, keywords, synonyms, and related
        topics. Consider multiple angles of inquiry.
  1.3 Verification Checklist:
      - Create a checklist based on the query's constraints. This
        guides the process and is used for final verification.

Phase 2: Iterative Research & Discovery
  Tool Usage:
      - search: Broad discovery and initial snippets.
      - open_page: Mandatory follow-up for promising results.
        Snippets are insufficient; analyze the full source.
  Query Strategy:
      - Start broad, narrow as you learn more.
      - Never repeat the exact same query. Rephrase or change angle.
      - Execute 5-50 tool calls depending on complexity.
  Post-Action Analysis: After every tool call, summarize findings,
  extract facts, and state how this affects your next step.

  Execute research using an iterative OODA loop:
    2.1 Observe: Review gathered information. Identify gaps.
    2.2 Orient:  Analyze effectiveness. Refine understanding.
    2.3 Decide:  Choose the most effective next action.
    2.4 Act:     Execute using available tools. Return to Observe.

Phase 3: Synthesis & Analysis
  3.1 Continuously integrate new information with existing knowledge.
  3.2 Triangulate critical data across 2+ independent sources.
  3.3 Handle dead ends: broaden scope, try alternative keywords,
      or research related context to uncover new leads.
  3.4 Maintain a running "Fact Sheet" of key facts and sources.

Phase 4: Verification & Final Report
  4.1 Review your Verification Checklist. Confirm sufficient evidence
      for each item from documents you have opened.
  4.2 If any item is unconfirmed, return to Phase 2 immediately.
  4.3 Never give up. Try different entities, broader context, related
      facts, or indirect routes. Every checklist item must reach
      [VERIFIED] or be explicitly ruled out with documented evidence.
  4.4 Construct the Final Report:
      - Synthesize all facts into a comprehensive answer.
      - Directly answer the original query.
      - Ensure all claims are supported by conducted research.
      - Submit using the finish tool with fields:
        Exact Answer, Explanation (cite by docid [N]), Confidence.
\end{lstlisting}

\end{document}